\documentclass[twocolumn,showpacs,preprintnumbers,aps]{revtex4}
\begin{document}

\title{A geometry for the electroweak field}
\author{Peter Morgan}
\email{peter.morgan@philosophy.oxford.ac.uk}
\affiliation{30, Shelley Road, Oxford, OX4 3EB, England.}
\homepage{http://users.ox.ac.uk/~sfop0045}
\altaffiliation{\\ Mail address until August 2003: 
      207 von Neumann Drive, Princeton NJ08540, USA.}

\date{\today}
\begin{abstract}
The structure of the electroweak theory is suggested by classical
geometrical ideas. A nonlinear map is constructed,
from a 12-dimensional linear space of three Weyl spinors onto the
12-dimensional tangent bundle of the Stiefel manifold of orthonormal
tetrads associated with the Lorentz group --- except, inevitably, for a
set of measure zero. In the approach of this paper, the electroweak field
is more natural than the Dirac field. This may be just a curiosity since it
may not survive quantization, but it suggests a path to bosonization of
the electroweak field in (3+1) dimensions.
\end{abstract}

\pacs{12.10.Dm,11.10.Lm}
\maketitle

\newcommand\RR{{\mathrm{I\hspace{-.2em}R}}}
\newcommand\HH{{\mathrm{H\hspace{-.65em}H}}}
\newcommand\ZZ{{Z\hspace{-.45em}Z}}
\newcommand\LL{{\sqrt{(\overline{\psi}\psi)^2+(\overline{\psi}\gamma^{0123}\psi)^2}}}
\newcommand\TZ {{T\hspace{-.35em}Z}}

This paper suggests that a classical, pre-quantization version of the
electroweak dynamics can be understood as a nonlinear dynamics
for an orthonormal tetrad field model. We will take as a starting point
the construction from a classical Dirac field of an orthonormal tetrad
field and its rate of change along the time-like 4-vector of the tetrad:
\begin{eqnarray*}
    T^\mu_{(0)}&=&\frac{\overline{\psi}\gamma^\mu\psi}{\LL},\cr
    T^\mu_{(1)}&=&\frac{\mathrm{Re}[\overline{\psi^c}\gamma^\mu\psi]}{\LL},\cr
    T^\mu_{(2)}&=&\frac{\mathrm{Im}[\overline{\psi^c}\gamma^\mu\psi]}{\LL},\cr
    T^\mu_{(3)}&=&\frac{\overline{\psi}i\gamma^{0123}\gamma^\mu\psi}{\LL},\cr
    S^{\mu\nu\alpha}T_{\alpha(0)}&=&S^{\mu\nu}=\overline{\psi}i\gamma^{\mu\nu}\psi,
\end{eqnarray*}
assuming that $\LL$ is non-zero almost everywhere, so that we can take
$T^\mu_{(i)}$ to be defined by continuity wherever $\LL$ is zero.
{[Notation: $\gamma^\mu$ is a generating basis for the Dirac algebra (which is a Clifford
algebra over the complex numbers), satisfying
$\gamma^\mu\gamma^\nu+\gamma^\nu\gamma^\mu=2g^{\mu\nu}$, where the metric
has signature $(1,-1,-1,-1)$;
$\gamma^{\mu\nu}=\frac{1}{2}(\gamma^\mu\gamma^\nu-\gamma^\nu\gamma^\mu)$;
$\gamma^{0123}=\gamma^0\gamma^1\gamma^2\gamma^3$.]}
$T^\mu_{(i)}$ and $S^{\mu\nu}$ given on a time-like hypersurface are sufficient
to determine Cauchy initial conditions for a second order differential equation for
$T^\mu_{(i)}$ in a manifestly Lorentz covariant way (since the values given at
a point are independent of which time-like hypersurface is chosen).
It is well-known that $T^\mu_{(0)}$ is time-like and orthogonal to $T^\mu_{(3)}$, but not
as well-known that taken with $T^\mu_{(1)}$ and $T^\mu_{(2)}$, constructed using
the charge conjugate of $\psi$, which is determined only up to a complex phase, they
form an orthonormal tetrad. As a result of Fierz identities, the six degrees of freedom
of $T^\mu_{(i)}$ and the six degrees of freedom of $S^{\mu\nu}$ are not independent
--- inevitably, since they derive from the eight degrees of freedom of $\psi$. The
orthonormal tetrad $T^\mu_{(i)}$ leaves the scale and relative phase
$\overline{\psi_R}\psi_L$ of the left and right components of $\psi$ undetermined.
A first order classical dynamics for $\psi$ can be understood to be equivalent to a
constrained higher order dynamics for the tetrad $T^\mu_{(i)}$.

To represent the six degrees of freedom of $T^\mu_{(i)}$ and the six degrees of
freedom of $S^{\mu\nu}$ independently using spinors, we will follow the structure
of the pre-quantization electroweak field, introducing two left-handed Weyl spinor
fields, a right-handed Weyl spinor field, and a Higgs field:
$$\Psi_L=\left(\begin{array}{c}\psi_{1L}\\ \psi_{2L}\end{array}\right),\qquad \psi_R,\qquad
     \Phi=\left(\begin{array}{c}\phi_1\\ \phi_2\end{array}\right).$$
Then, we can construct two Dirac fields,
$$\psi=\Phi^\dagger\Psi_L+\psi_R,\qquad\chi=\Phi^{c\dagger}\Psi_L+\psi_R,$$
where $\Phi^c$ is a conjugate field orthogonal to $\Phi$ and of the same scale,
which is again determined only up to a complex phase. From these we can construct
an orthonormal tetrad field as above,
\begin{eqnarray*}
    T^\mu_{(0)}&=&\frac{\overline{\psi}\gamma^\mu\psi}{\LL},\cr
    T^\mu_{(1)}&=&\frac{\mathrm{Re}[\overline{\psi^c}\gamma^\mu\psi]}{\LL},\cr
    T^\mu_{(2)}&=&\frac{\mathrm{Im}[\overline{\psi^c}\gamma^\mu\psi]}{\LL},\cr
    T^\mu_{(3)}&=&\frac{\overline{\psi}i\gamma^{0123}\gamma^\mu\psi}{\LL},
\end{eqnarray*}
and construct an independent antisymmetric tensor $S^{\mu\nu}$ for the rate of change
of $T^\mu_{(i)}$ along the time-like 4-vector $T^\mu_{(0)}$ as a linear combination
of $\overline{\chi}i\gamma^{\mu\nu}\chi$ and
$\overline{\chi^c}\gamma^{0123}\gamma^{\mu\nu}\chi$
that is determined by the scale and relative phase
$\xi=\overline{\psi_R}\Phi^\dagger\Psi_L$, 
$$S^{\mu\nu\alpha}T_{\alpha(0)}=S^{\mu\nu}=
          |\xi|^2\overline{\chi}i\gamma^{\mu\nu}\chi+
          \mathrm{Re}[\xi\overline{\chi^c}\gamma^{0123}\gamma^{\mu\nu}\chi].$$
With this construction, $T^\mu_{(i)}$ and $S^{\mu\nu}$ are entirely independent. This is
a little surprising, since $\Phi$ has played no essential r\^ole: $\Phi^\dagger\Psi_L$
and $\Phi^{c\dagger}\Psi_L$ could have been written just as two left handed spinors, so that
a 12-dimensional linear space is mapped nonlinearly and covariantly onto the 12-dimensional
tangent bundle of the Stiefel manifold of orthonormal tetrads associated with the Lorentz group
--- except, inevitably, for a set of measure zero.

The algebra and geometry of the above deserve a little expansion. For the algebra,
by introducing the charge conjugate Dirac spinor into the algebra, we effectively make the
Dirac representation into a real matrix representation
$M_8(\RR)\simeq M_2(\HH_1)\otimes \HH_2$
of the quaternionic Clifford algebra that can be constructed from the Lorentzian metric.
In terms of this algebraic structure, the ``j'' and ``k'' basis elements of the quaternionic
algebra $\HH_2$, together with the complex ``i'' considered as a basis element of
$\HH_2$, give a system of bilinear constructions parallel to the constructions
$\overline{\psi}i\gamma^{0123}\gamma^\mu\psi$ and $\overline{\psi}i\gamma^{\mu\nu}\psi$ for
a Dirac spinor $\psi$, and related by an $SU(2)$ group action that generalizes the $U(1)$
group action of the Dirac algebra considered without charge conjugation. The action of
this $SU(2)$ group leaves $\overline{\psi}\gamma^\mu\psi$ invariant, so it can be
considered to be the little group of $\overline{\psi}\gamma^\mu\psi$, of Euclidean rotations
acting on the Lorentz covariant bilinear forms
$\overline{\psi}i\gamma^{0123}\gamma^\mu\psi$ and $\overline{\psi}i\gamma^{\mu\nu}\psi$.
We could write a basis for the full set of bilinear forms in a quaternionic notation as
$\overline{\psi}\psi$,
$\overline{\psi}\gamma^\mu\psi$, 
$\mathrm{Re}[\overline{\psi}i\gamma^{\mu\nu}\psi]$,
$\mathrm{Re}[\overline{\psi}j\gamma^{\mu\nu}\psi]$,
$\mathrm{Re}[\overline{\psi}k\gamma^{\mu\nu}\psi]$,
$\mathrm{Re}[\overline{\psi}i\gamma^{0123}\gamma^\mu\psi]$,
$\mathrm{Re}[\overline{\psi}j\gamma^{0123}\gamma^\mu\psi]$,
$\mathrm{Re}[\overline{\psi}k\gamma^{0123}\gamma^\mu\psi]$, and
$\overline{\psi}\gamma^{0123}\psi$, if the Dirac conjugate $\overline{\psi}$
is extended to act as a quaternionic conjugation on $i$, $j$, and $k$.

For the geometry, $\chi$ does not give us a sufficiently free choice to allow us to set
$S^{\mu\nu}=\overline{\chi}i\gamma^{\mu\nu}\chi$, because of the constraint on
$\psi_R$ imposed by the construction of $T^\mu_{(i)}$. Recall that a given
antisymmetric tensor $S^{\mu\nu}$ can be written as
$\kappa A^{\mu\nu}+\lambda B^{\mu\nu}$, where $A^{\mu\nu}=T^{[\mu}Z^{\nu]}$, with
$T^\mu T_\mu=1$, $Z^\mu Z_\mu=-1$, $T^\mu Z_\mu=0$, $A^{\mu\nu}B_{\mu\nu}=0$;
let $\TZ$ be the 2-space spanned by $T^\mu$ and $Z^\mu$.
If we choose $\overline{\chi}\gamma^\mu\chi\in \TZ$, as we can, the scale and phase of
$\xi$ are fixed by the direction in the 3-space $V_\chi$ --- spanned by
$\mathrm{Re}[\overline{\psi}i\gamma^{0123}\gamma^\mu\psi]$,
$\mathrm{Re}[\overline{\psi}j\gamma^{0123}\gamma^\mu\psi]$, and
$\mathrm{Re}[\overline{\psi}k\gamma^{0123}\gamma^\mu\psi]$ ---
of the intersection $Z_\chi$ of $\TZ$ with $V_\chi$. $\xi$ is effectively used to construct a
stereographic projection into $V_\chi$. $\psi_R$ is fixed by $\xi$, and $\chi_L$ is
fixed by the choice of $\overline{\chi}\gamma^\mu\chi$ up to a phase, which is fixed by the
phase of $(\kappa,\lambda)$. With $\psi_R$ and $\chi_L$ fixed, the scale
of $S^{\mu\nu}_{(\xi,\chi)}=|\xi|^2\overline{\chi}i\gamma^{\mu\nu}\chi+
          \mathrm{Re}[\xi\overline{\chi^c}\gamma^{0123}\gamma^{\mu\nu}\chi]$
is fixed, which almost certainly will not be the scale required to give
$S^{\mu\nu}=S^{\mu\nu}_{(\xi,\chi)}$, although we \emph{have} succeeded in constructing
$S^{\mu\nu}_{(\xi,\chi)}$ to be proportional to $S^{\mu\nu}$ --- so we will have to choose
$\overline{\chi}\gamma^\mu\chi\in \TZ$ carefully (unsurprisingly, since we are mapping
between two manifolds of the same dimension).
Changing our choice of $\overline{\chi}\gamma^\mu\chi\in \TZ$ generally changes the
direction of the intersection $Z_\chi$ in $V_\chi$; provided $\TZ$ does not contain
$\overline{\chi}i\gamma^{0123}\gamma^\mu\chi$, we can vary the direction of $Z_\chi$
relative to $\overline{\chi}i\gamma^{0123}\gamma^\mu\chi$ continuously between almost
parallel and almost anti-parallel, thereby changing the scale of $\xi$ continuously in the
range $(0,\infty)$, and hence changing the scale of $S^{\mu\nu}_{(\xi,\chi)}$. This is
enough to ensure that, except for a set of measure zero, we can construct a representation
in terms of three Weyl spinors for any given $T^\mu_{(i)}$ and $S^{\mu\nu}$.

It is the suggestion of this paper that the dynamics of the electroweak
field has its particular form because it describes a perturbation of a
second order dynamics of an orthonormal tetrad field. To satisfactorily do so, the
symmetries of the algebraic equations used to construct $\psi$ and $\chi$
under the action of $SU(2)$, $\Phi=U\Phi$, $\Psi_L=U\Psi_L$, and under
the action of scaling, $\Phi=\alpha\Phi$, $\Psi_L=\alpha^{-1}\Psi_L$,
$\alpha\in\RR^+$, should also be symmetries of the dynamics of the
electroweak field. The $SU(2)$ symmetry is of course present in the dynamics
of the electroweak theory, but the scaling symmetry is conspicuously absent,
which ultimately would suggest that the Higgs particle might not be found, since it
corresponds to the scale degree of freedom of the Higgs field.

The nonuniqueness of the construction above consists in a choice of an $SU(2)$ phase
for $\Phi$, but also consists in a fixed choice of an $O(3)$ orientation and phase for the
assignment of $T^\mu_{1}$, $T^\mu_{2}$, and $T^\mu_{3}$ to orthogonal linear
combinations of
$\mathrm{Re}[\overline{\psi}i\gamma^{0123}\gamma^\mu\psi]$,
$\mathrm{Re}[\overline{\psi}j\gamma^{0123}\gamma^\mu\psi]$, and
$\mathrm{Re}[\overline{\psi}k\gamma^{0123}\gamma^\mu\psi]$,
and finally in a fixed choice of an $O(3)$ orientation and phase for the stereographic
projection of $\xi$ into $V_\chi$.
The nonuniqueness of the construction above does not affect the suggestion
above. A different construction just results in a given dynamics of the electroweak field
inducing a different dynamics of the orthonormal tetrad field that we construct.
Perhaps no dynamics of an orthonormal tetrad field derived from a dynamics of an
electroweak field will be simple to express, but a particular construction may, or may not,
be picked out by the relative simplicity of equations describing the induced dynamics.
The suggestion above is determined by no more than the Lorentzian structure of
spinors, the number of degrees of freedom and the symmetries of the dynamics
of the electroweak field, and the dimensionality of the Lorentz group. The particular
representation above, of the Lorentz group acting on an orthonormal tetrad field,
should not be considered essential.

If we do think of the dynamics of the electroweak field as inducing a dynamics
of an orthonormal tetrad field, then there is a topological invariant, corresponding
to the homotopy class of the orthonormal tetrad field. The homotopy group for
maps from $S^3$ to the Lorentz group (supposing that the dynamics effectively
compactifies $\RR^3$ to $S^3$) is just \ZZ. Following the Mandelstam approach
to bosonization\cite{Mandelstam}, fermion operators appear as soliton operators
in the framework of a bosonic theory, and fermion charges appear as topological
charges of the bosonic theory\cite{Marino}. 
For a quantized sine-Gordon field $\hat\phi(y)$ in (1+1) dimensions, Mandelstam
constructs two operators $\hat\psi_1(x)$ and $\hat\psi_2(x)$, with commutation
relations, at a fixed time $t$,
$$[\hat\phi(y),\hat\psi_i(x)]=2\pi\beta^{-1}\theta(x-y)\hat\psi_i(x);$$
$\hat\psi_1(x)$ and $\hat\psi_2(x)$ also satisfy the anticommutation relations
and dynamics of the 1+1 dimensional Dirac equation. $\hat\psi_1(x)$ and
$\hat\psi_2(x)$ change the homotopy class of the sine-Gordon field
$\hat\phi(y)$, supposing that the sine-Gordon field is effectively a map
from $S^1$ to $S^1$. Mandelstam's approach suggests
$S^3$ as a topologically nontrivial starting point for a bosonization of the Dirac
equation in 1+3 dimensions, but the numbers of degrees of freedom in the 1+3
dimensional case do not match, as they do in the 1+1 dimensional case ---
$3\times 2\ne 8$ degrees of freedom, in contrast to $1\times 2=2$. $S^3$ might
nonetheless work for the 1+3 dimensional Dirac equation (sadly, there isn't a general
theory that can assure us that we can in principle bosonize any given fermion field),
but, at least in the light of the construction given above, $SO(1,3)$ is the natural
structure to try as the basis of a bosonization of the electroweak theory.

The topological invariant of the orthonormal tetrad field can also be thought of as a
topological invariant of the first $O(3)$ nonuniqueness discussed in the last but one
paragraph, if we take that $O(3)$ orientation and phase to be variable. There is a
curious further possibility for us to introduce raising and lowering operators for the
second $O(3)$ nonuniqueness, if we take that $O(3)$ orientation and phase also
to be variable. All of this may, \emph{perhaps}, allow an understanding of the
relationship between the three families of leptons.

The inverse square roots don't look good from the point of view of quantization. They
can be removed, but only at the cost of diluting the topological properties of the
construction above. We can construct two Dirac fields as above,
$\psi=\Phi^\dagger\Psi_L+\psi_R$, $\chi=\Phi^{c\dagger}\Psi_L+\psi_R$,
then we can construct an orthogonal tetrad field,
\begin{eqnarray*}
    T^\mu_{(0)}&=&\Phi^\dagger\Phi\overline{\psi}\gamma^\mu\psi,\cr
    T^\mu_{(1)}&=&\Phi^\dagger\Phi\mathrm{Re}[\overline{\psi^c}\gamma^\mu\psi],\cr
    T^\mu_{(2)}&=&\Phi^\dagger\Phi\mathrm{Im}[\overline{\psi^c}\gamma^\mu\psi],\cr
    T^\mu_{(3)}&=&\Phi^\dagger\Phi\overline{\psi}i\gamma^{0123}\gamma^\mu\psi;
\end{eqnarray*}
for each of these 4-vectors the length $\left |T^\mu_{(i)}\right|$ is $\Phi^\dagger\Phi\LL$.
We again define $\xi=\overline{\psi_R}\Phi^\dagger\Psi_L$, then we can construct a
tensor $S^{\mu\nu}$ for the rate of change of $T^\mu_{(i)}$ along the time-like 4-vector
$T^\mu_{(0)}$ as
$$S^{\mu\nu}= |\xi|^2\overline{\chi}i\gamma^{\mu\nu}\chi+
                \mathrm{Re}[\xi\overline{\chi^c}\gamma^{0123}\gamma^{\mu\nu}\chi] +
                    T^\beta_{(0)}\partial_\beta(\Phi^\dagger\Phi) g^{\mu\nu}.$$
We can use the rate of change of $\Phi^\dagger\Phi$ along $T^\mu_{(0)}$ for the
symmetric part of $S^{\mu\nu}$ because $\Phi$ satisfies a second order differential
equation in the electroweak theory. This construction, which is again not unique, restores
an expectation that we will find the Higgs particle.

The electroweak theory is a rather ugly construction, with its three Weyl spinors
and a Higgs field. It is empirically well justified, but it has very little theoretical
motivation. The construction of this paper seems a moderately natural
parameterization of the tangent bundle of the Lorentz group, which may
be construed as a weak theoretical motivation for the electroweak theory.

I am grateful to Jonathan Dimock for comments on earlier versions of this paper.


\begin{thebibliography}{Mandelstam}
\bibitem{Mandelstam}
  {S. Mandelstam, Phys. Rev. D \textbf{11}, 3026 (1975).}
\bibitem{Marino}
  {E.C.Marino, Phys. Lett. B \textbf{263}, 63 (1991)}
\end{thebibliography}
\end{document}